\documentclass[final,1p,times]{elsarticle} 
\usepackage{graphicx} 
\usepackage{amssymb} 
\usepackage{amsthm} 
\usepackage{lineno} 

\journal{Nuclear Physics A} 
\begin{document} 

\begin{frontmatter} 


\title{QCD Phase Diagram: Phase Transition, Critical Point and Fluctuations}

\author{Bedangadas Mohanty$^{a}$}

\address[a]{Variable Energy Cyclotron Centre, 
1/AF, Bidhan Nagar,
Kolkata, 700064, India}

\begin{abstract} 
A summary of discussions on selected topics related to QCD phase diagram, phase transition, 
critical point, fluctuations and correlations at the Quark Matter 2009 conference are presented.
\end{abstract} 

\end{frontmatter} 



\section{Introduction}

Understanding the properties of the nuclear matter when subjected to extremes of 
temperature and density is one of the goals of the relativistic heavy-ion collision 
program. The knowledge of the properties manifested in these collisions should have 
some connection to our understanding of the early universe and how hadrons acquire their masses. 
QCD suggests we should at least expect two types of transitions for nuclear matter in limits
of high temperature ($T$) and densities : deconfinement transition and chiral phase transition. 
Theoretically the deconfinement measure is the order parameter, 
Polyakov loop ($L(T) \sim \lim_{r \to \infty}~\exp\{-V(r)/T\}$,
where $V(r)$ is the potential between a static quark-antiquark pair
separated by a distance $r$). The measure of chiral transition is the 
order parameter, the chiral condensate $\langle {\bar \psi} \psi \rangle (T)$.
For a confined phase for $T < T_c$ (critical temperature for transition)
we have  $L(T) \simeq 0$ and $\langle {\bar \psi} \psi \rangle(T) \not= 0$, and for 
a deconfined phase for $T>T_c$ with $L(T)\not= 0$ 
and $\langle {\bar \psi} \psi \rangle(T) \simeq 0$. 
In this proceedings we summarize the theoretical and experimental 
understanding of the QCD phase diagram as was discussed at the QM2009.

\section{Lattice QCD results}

Lattice calculations simulate a quantum statistical ensemble in
thermal equilibrium at fixed $T$ with partition function
$Z = Tr~{\rm e}^{-H/T}$, where H is the QCD Hamiltonian. The lattice spacing
$a$ and $N_\tau$ sites in imaginary time are related to $T$ as $aN_t = 1/T$.
Usually $a$ is varied to change $T$ keeping $N_\tau$ fixed. An approach 
based on ``$T$ integral method'' was presented, where $N_\tau$ is
varied and $a$ fixed to change $T$~\cite{tintegral}. There are few aspects 
to be noted while interpreting lattice results. To connect with reality, 
lattice results needs to be presented as extrapolations to continuum limit. The calculations 
are now done on lines of constant physics, along which the variations in 
observables can be attributed to changes in temperature and not also to 
changes in the Hamiltonian. Choice of proper actions (see recent work~\cite{action}), 
spatial volume and setting of quark masses are important for interpreting the results. 
Results are also dependent on the choice of number of quark flavours. The current view is, 
computation with 2 light quarks and a heavy strange quark (2+1) are close to a
realistic picture. At the conference, several studies from Lattice QCD 
were presented. These provided insights on the order of phase transition, transition 
temperature ($T_{c}$) , Equation-Of-State (EOS) and QCD critical point (QCP), some of these 
are discussed below.

\subsection{Order of phase transition}
\begin{figure}
\begin{minipage}[t]{0.3\textwidth}
  \includegraphics[width=1.02\textwidth]{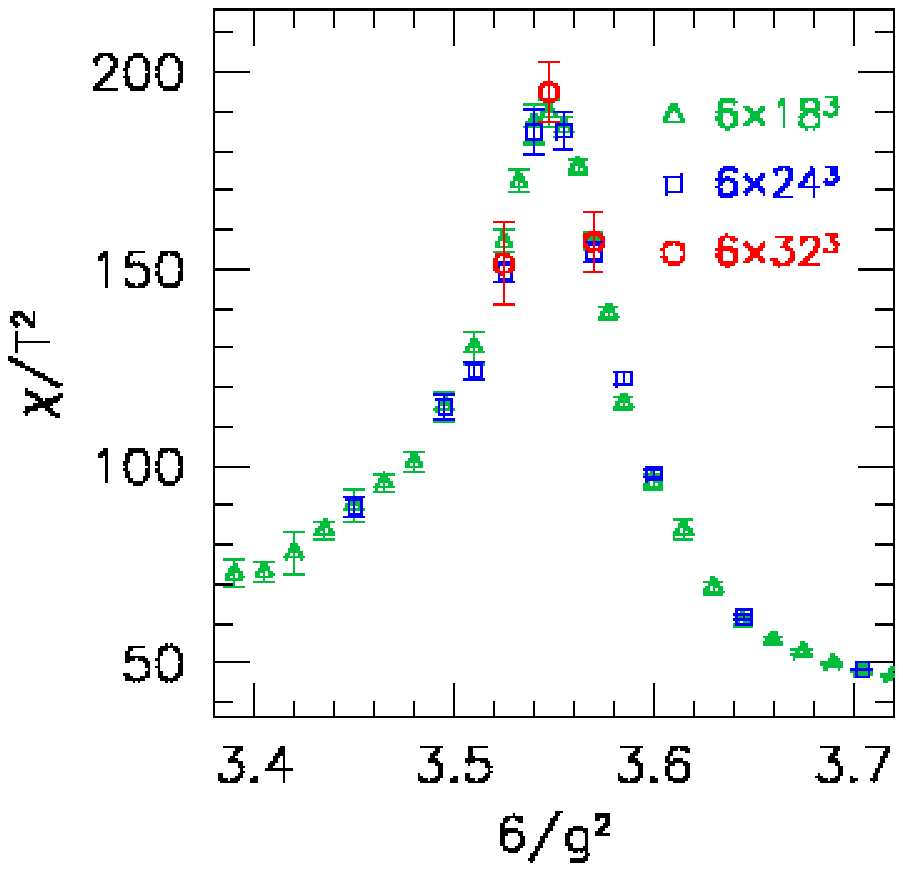}
\end{minipage}
\hspace{0.0\textwidth}
\begin{minipage}[t]{0.3\textwidth}
  \includegraphics[width=1.02\textwidth]{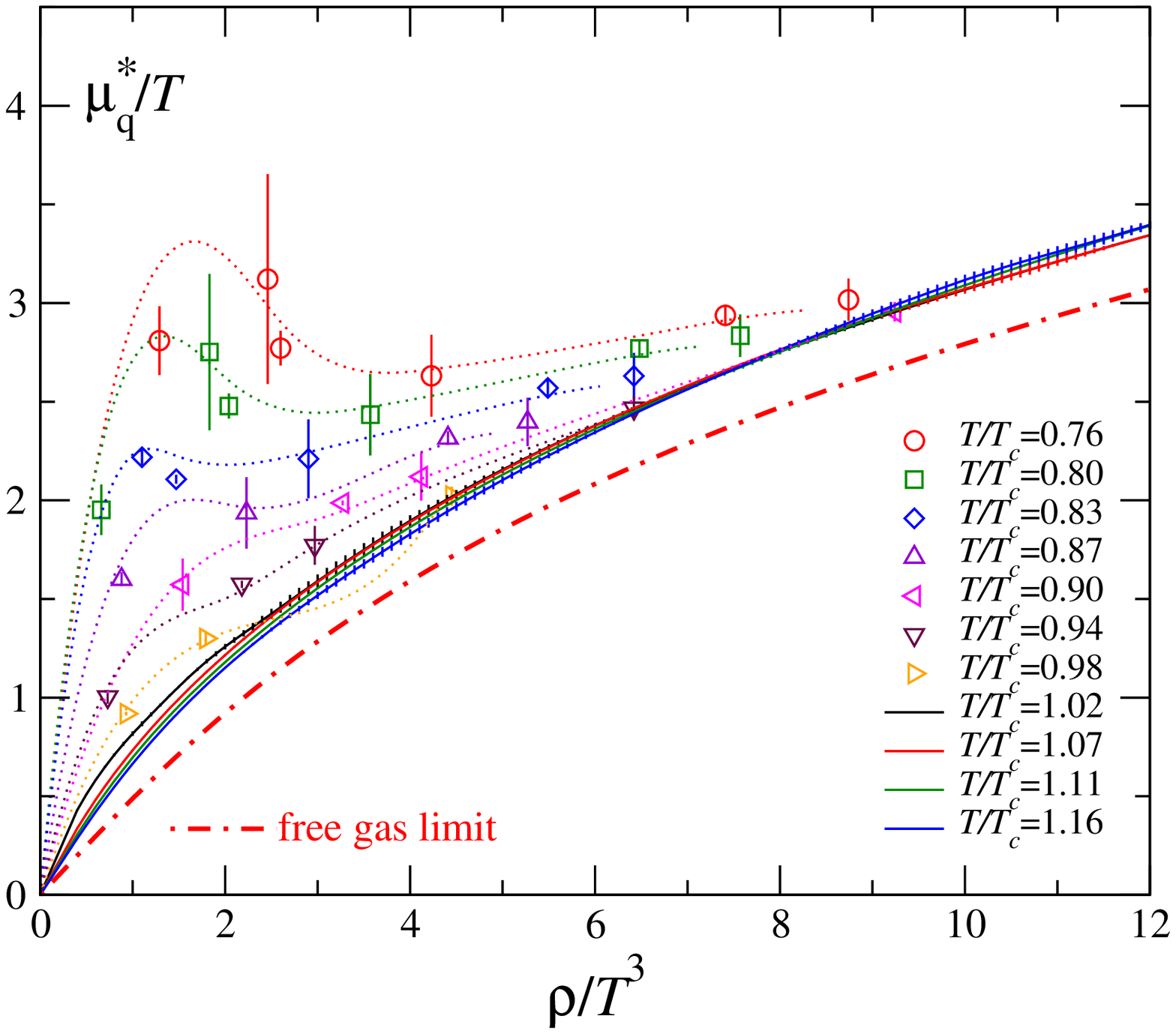}
\end{minipage}
\hspace{0.0\textwidth}
\begin{minipage}[b]{0.45\textwidth}
\caption{Left: Susceptibility  for the light quarks for $N_\tau$=6 
as a function of $6/g^2$, where $g$ is the gauge coupling
($T$ grows with $6/g^2$)~\cite{crossover}. 
Right: Derivative of logarithm of partition function in canonical ensemble formalism 
as a function of the quark number density~\cite{firstorder}.}
\label{co}
\end{minipage}
\end{figure}

QCD calculations on lattice at high temperature and $\mu_{B}$ = 0 MeV has established
the quark-hadron transition to be a cross-over~\cite{crossover}. Figure~\ref{co}
shows the lattice chiral susceptibility 
$\chi(N_s,N_t)$ = $\partial^2$/$(\partial$$m_{ud}^2)$($T/V$)$\cdot\log Z$,
where $m_{ud}$ is the mass of the light u,d quarks, $N_s$ is the spatial
extension, $N_\tau$ euclidean time extension,  and 
V the system volume. The susceptibility plotted as a function of 6/$g^2$ (g is the 
gauge coupling and $T$ grows with  6/$g^2$) shows a pronounced peak around the 
$T_c$. The peak and width are independent of volume
(varied by a factor 8) thereby establishing the transition to be an analytic 
cross-over~\cite{crossover}.
For a first-order phase transition
the height of the susceptibility peak should have been $\propto$ $V$ and
the width of the peak $\propto$ 1/$V$, while for a second-order transition the
singular behaviour should have been $\propto$ $V^\alpha$, $\alpha$ is a 
critical exponent.  A lattice result discussing the nature of phase transition 
at finite $T$ and baryon density~\cite{firstorder} was presented. It was based on 
idea that the chemical potential ($\mu_{q}^{*}$) which gives a minimum of the effective potential
does not increase when it passes through the mixed phase. Whereas for a cross-over transition
it shows a monotonic increase with density when $T$ $>$ $T_c$.
The results are shown in the right panel of Fig.~\ref{co}. This calculation which uses 
P4-improved staggered quark action, in canonical ensemble with $m_{\pi}$ = 700 MeV
and number of quark flavour of 2, results in a first order phase transition 
for $T/T_c$ $<$ 0.83 and $\mu_q/T$ $>$ 2.3~\cite{firstorder}.

\subsection{Transition Temperature}
The point of sharpest change in temperature dependence of the
chiral susceptibility ($\chi_{\bar{\psi}\psi}$), 
the strange quark number susceptibility ($\chi_s$) and the renormalized Polyakov-loop (L) are
used to estimate the QCD transition temperature in lattice calculations.
The recent results on chiral and deconfinement phase transition temperatures were presented by
two groups (HotQCD/RBC~\cite{rbc} and Budapest/Wuppertal~\cite{buda}).  
Using the observable $\chi_{\bar{\psi}\psi}/T^2$
the HotQCD/RBC and Budapest/Wuppertal groups get the chiral phase transition temperatures as
192 (4)(7) MeV and 152(3)(3) MeV respectively. Using the observable $L$ the deconfinement 
transition temperature remained the same for HotQCD/RBC and is 170(4)(3) MeV from 
Budapest/Wuppertal group.
The two groups differ in the transition temperatue for both chiral and de-confinement 
transitions. The possible sources of differences in the two approaches could be ambiguity in locating 
$T_c$ for a cross-over, physical observable used to set the scale, 
preferred renormalization of chiral susceptibilities and choice of actions. 
It was suggested at the conference to study the QCD thermodynamics with a theoretically 
firmly established Wilson type fermion discretization.
Such a large difference in the value of $T_{c}$ has serious consequences for heavy-ion 
phenomenological studies.
As energy density ($\epsilon$) $\sim$ $T^{4}$, the lack of accurate determination of $T_{c}$ leads to
about 60\% uncertainty in $\epsilon$ at $T_{c}$.

\subsection{Equation of State}
\begin{figure}
\begin{minipage}[t]{0.29\textwidth}
\includegraphics[width=1.04\textwidth]{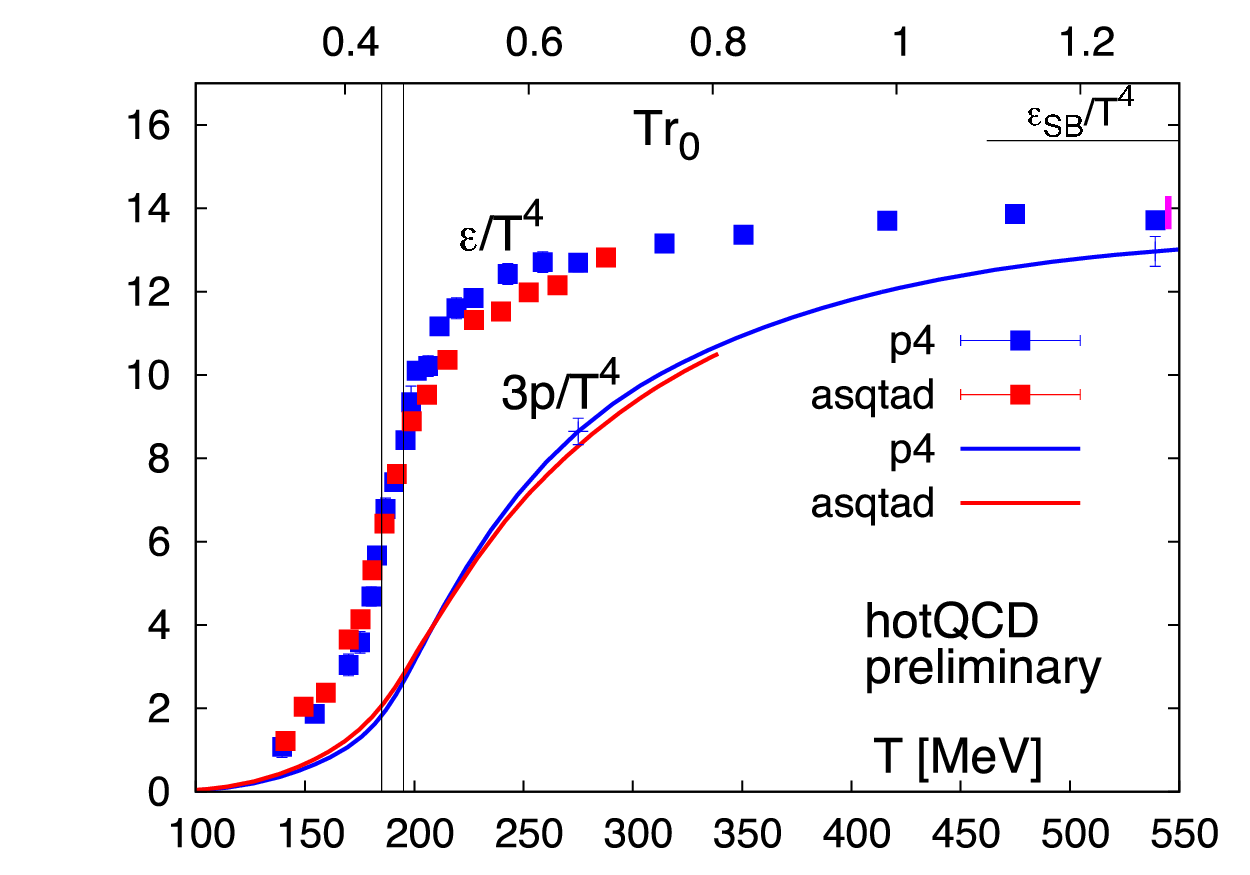}
\end{minipage}
\begin{minipage}[t]{0.3\textwidth}
\includegraphics[width=1.04\textwidth]{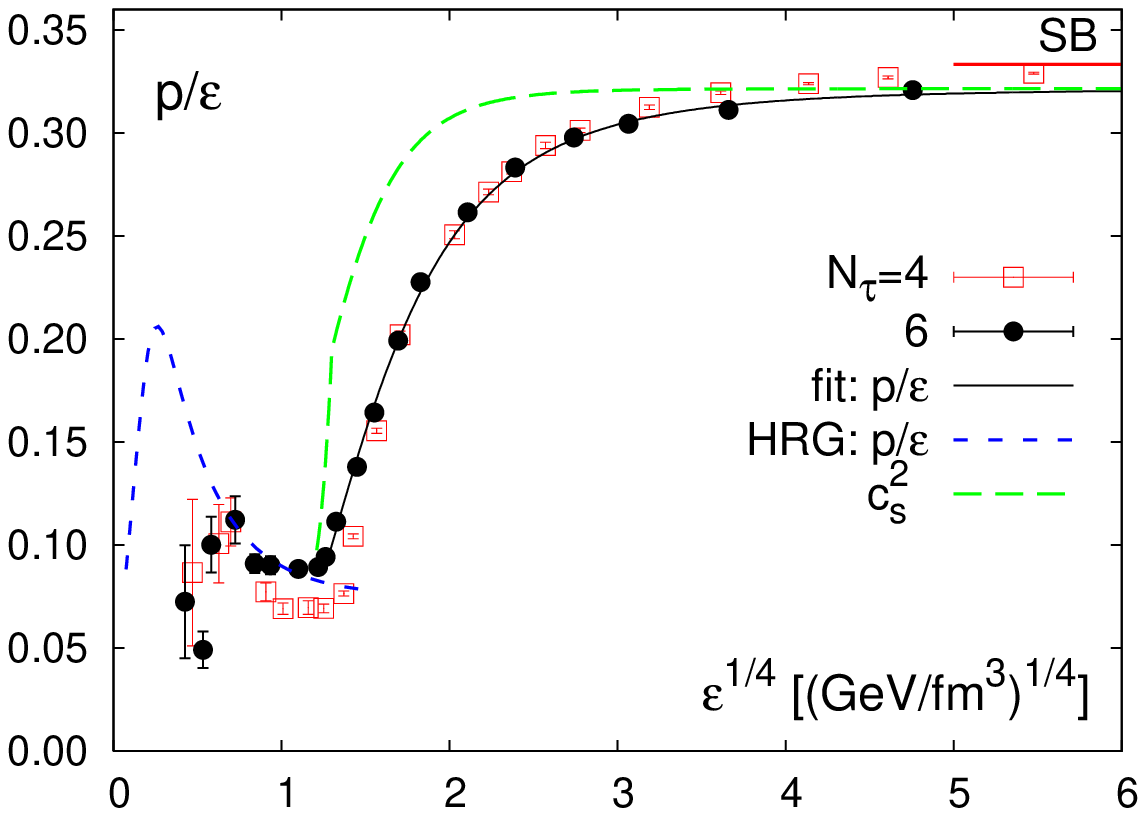}
\end{minipage}
\begin{minipage}[b]{0.35\textwidth}
 \caption{Left: Results for energy density ($\epsilon$) and pressure ($p$) 
for different actions. The band T = 185-195 MeV, drawn to guide the eye, covers the inflection point~\cite{rgupta}.
Right: The ratio $p/\epsilon$ as function of the fourth root of the energy density
obtained from calculations on lattice~\cite{cheng}.
\vspace*{3mm}
}
 \label{eqn}
\end{minipage}
\end{figure}

Figure~\ref{eqn} (left) shows the variation of $\epsilon$/$T^{4}$ vs. $T$~\cite{rgupta,cheng}. As 
$\epsilon$/$T^{4}$ $\propto$ effective degrees of freedom (dof) of the system ($g_{\mathrm {eff}}$), 
the trend shows a change in $g_{\mathrm {eff}}$ below and above $T_{c}$. This is 
interpreated as a transition from a state with hadronic dof to a state with quark-gluon dof.
At high $T$ the $\epsilon$/$T^{4}$ is about 10-15\% lower compared to Stefan-Boltzman (SB) limit
indicating that the matter still has strong interaction effects. The deviations from ideal gas 
behaviour could be understood in terms of effective thermal masses of quarks and gluons,
and a consequence of this sizable interaction at high $T$ could lead to the existence of coloured
resonance states. It was shown that at very high $T$ $\sim$ $10^{5}-10^{7}$ MeV the
lattice calculations approach the SB limit and has good agreement with calculations from 
perturbation theory~\cite{fodor2}.
The functional relation between pressure ($p$)
and $\epsilon$ is the EOS. The velocity of sound in 
the medium ($c_{s}^{2}$) is $d{p}/d{\epsilon}$ at fixed entropy. 
$p/\epsilon$ is plotted as a function of $\epsilon^{1/4}$ computed from Lattice in 
Fig.~\ref{eqn} (right)~\cite{cheng}. This is
an important ingredient to heavy-ion phenomenology calculations, as the $c_{s}$ decides 
the rate of cooling of the high-$T$ system and provides input on the system composition
at different evolution times. For the dependence of thermodynamics of QCD plasma 
on number of colors see~\cite{marco}.

\section{Experimental results on Charge Correlations}

\begin{figure}[tbp]
\begin{minipage}[t]{0.4\textwidth}
\centerline{ \includegraphics[width=0.96\textwidth]{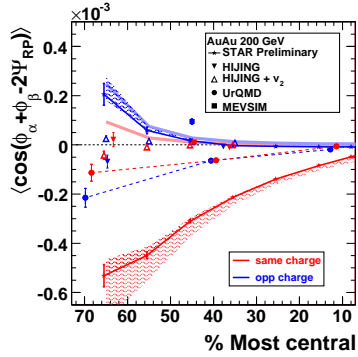}}
\end{minipage}
\begin{minipage}[b]{0.45\textwidth}
 \caption{Azimuthal charge correlations in data compared to simulation results 
for 200~GeV Au+Au~\cite{voloshin}.  Blue symbols mark  {opposite-charge}
  correlations, and red are  {same-charge}.
  Markers connected by solid lines  represent the data.
\vspace*{3mm}
}
 \label{parity}
\end{minipage}
\end{figure}

The STAR experiment at RHIC presented preliminary 
results on charged hadron azimuthal correlations based on
3-particle correlation technique~\cite{voloshin}. 
The results from Au+Au collisions at $\sqrt{s_{NN}}$ = 200 GeV at
midrapidity  for 0.15 $<$ $p_{T}$ $<$ 2 GeV/$c$ 
between same charged and opposite charged hadrons with respect to reaction plane
(plays the role of third particle) are shown in Fig.~\ref{parity}. 
The observable, $\langle \cos(\phi_a +\phi_\beta -2\psi_{RP}) \rangle$ 
represented the difference between azimuthal correlations projected
onto the direction of the angular momentum vector and correlations projected
onto the collision event plane. 
The difference between the same charge and opposite charge correlations 
could not be explained by models such as HIJING and UrQMD 
and by incorporating realistic values for the 
elliptic flow in such simulations.
The signal seems to be consistent with the predictions for
existence of metastable domains in QCD vacuum leading to local {\it Parity} violation
in the vicinity of deconfinement transition expected to be
achieved in heavy-ion collisions~\cite{kharzeev}. For such a phenomena where the massless 
quarks can change  their chirality due to interactions with gluon fields, 
there could be separation of positive charges from negative charges 
along the direction of angular momentum of the collision 
as a result of large magnetic fields reached in the collisions (especially in 
non-central collisions). Deconfinement allows for the possibility of quarks 
traveling over distances greater than nucleonic scales and chiral symmetry 
restoration is essential, because a chiral condensate will tend to erase any 
asymmetry between the number of right- and left-handed fermions. 
The observable presented is parity-even~\cite{voloshin}, making it succeptible to physical
processes not related to parity violation effects. Thus understanding
of the physical background is crucial. The signal was also observed to be not
restricted in the low $p_{T}$ region as naively might be expected for parity-violation 
effects. 

\section{QCD Critical Point}

\subsection{Theory Calculations }

\begin{figure}[tbp]
\begin{minipage}[t]{0.3\textwidth}
\centerline{ \includegraphics[width=1.04\textwidth]{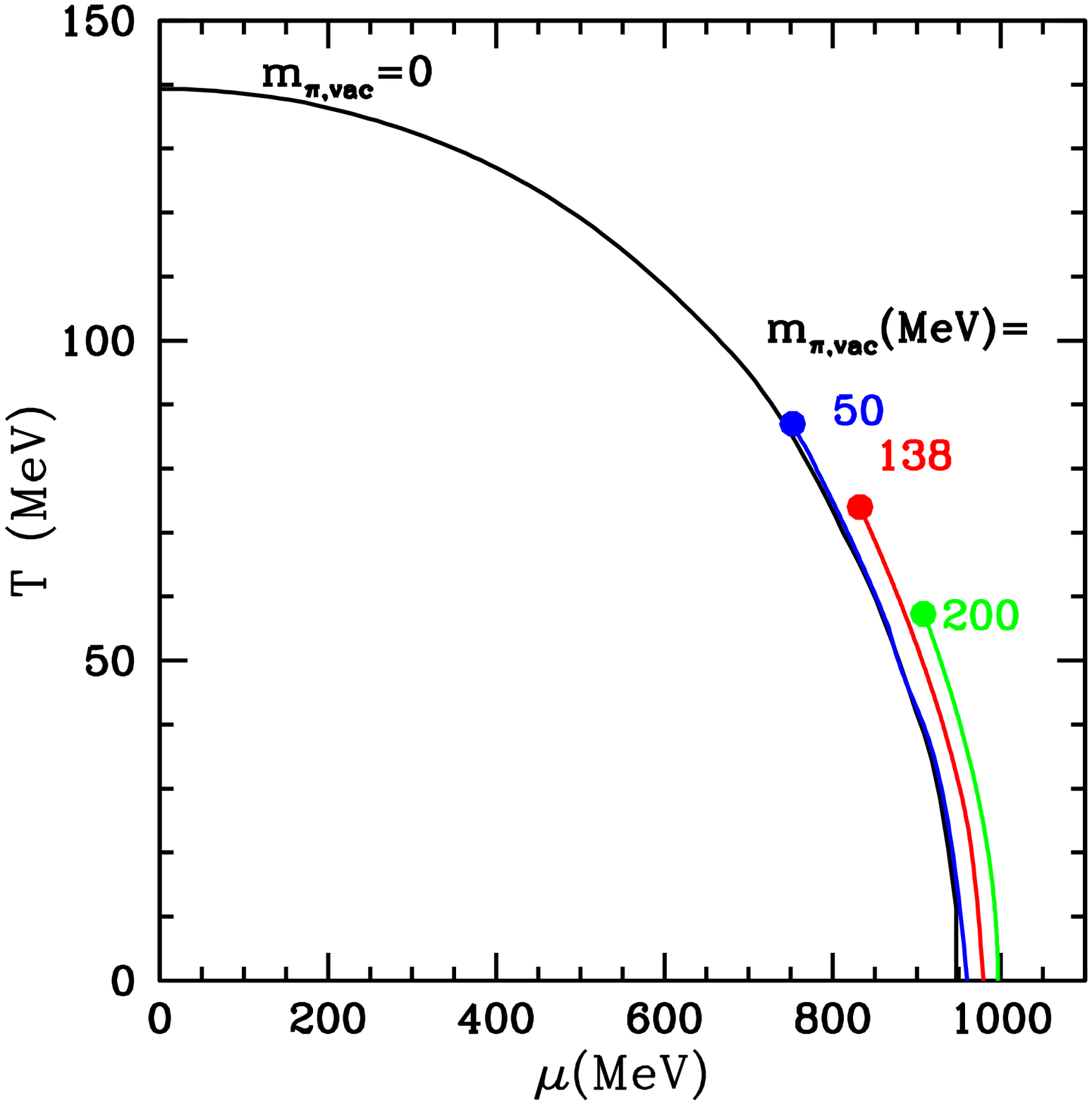}}
\end{minipage}
\begin{minipage}[t]{0.35\textwidth}
\centerline{ \includegraphics[width=1.04\textwidth]{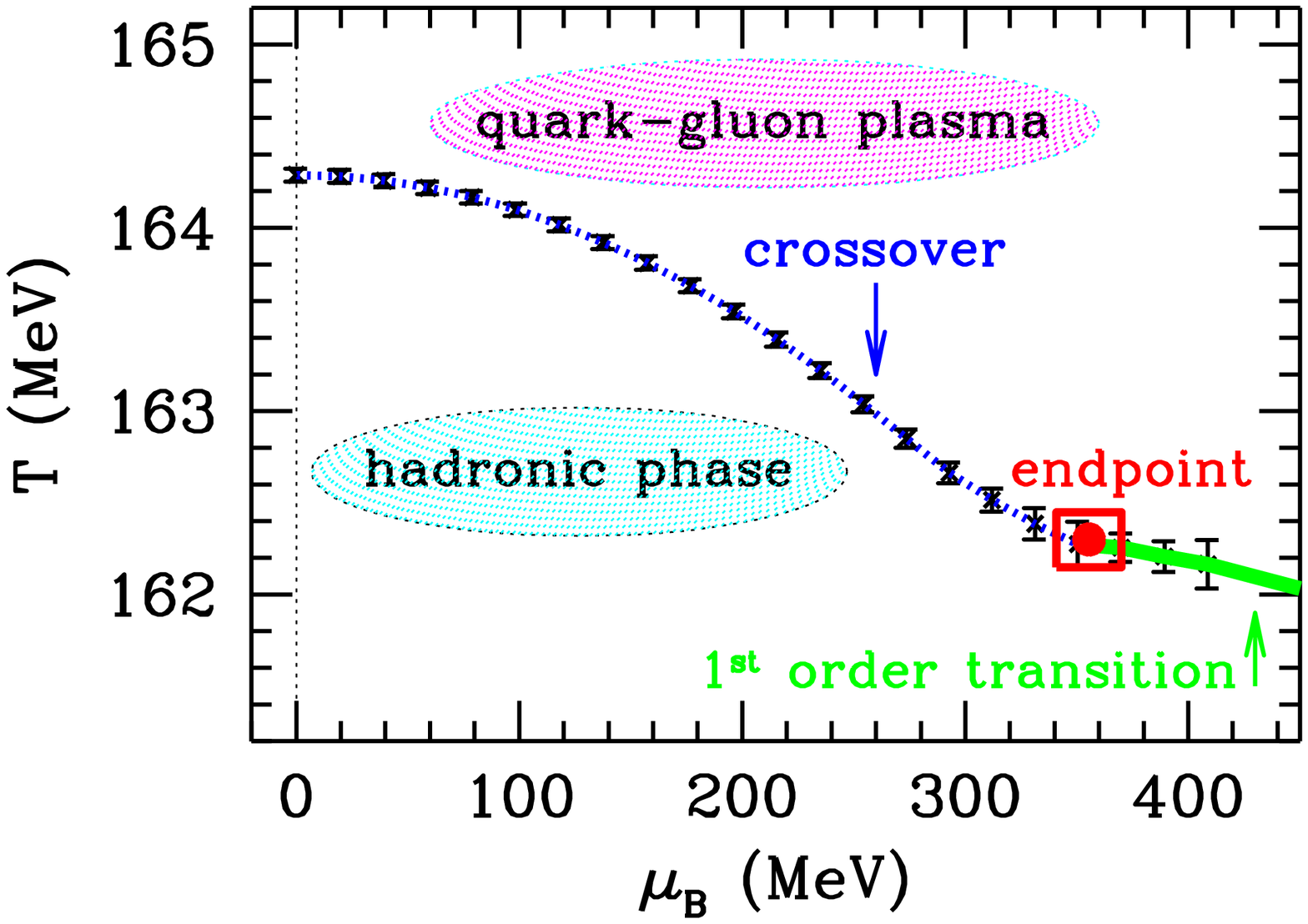}}
\end{minipage}
\begin{minipage}[t]{0.34\textwidth}
\centerline{ \includegraphics[width=1.04\textwidth]{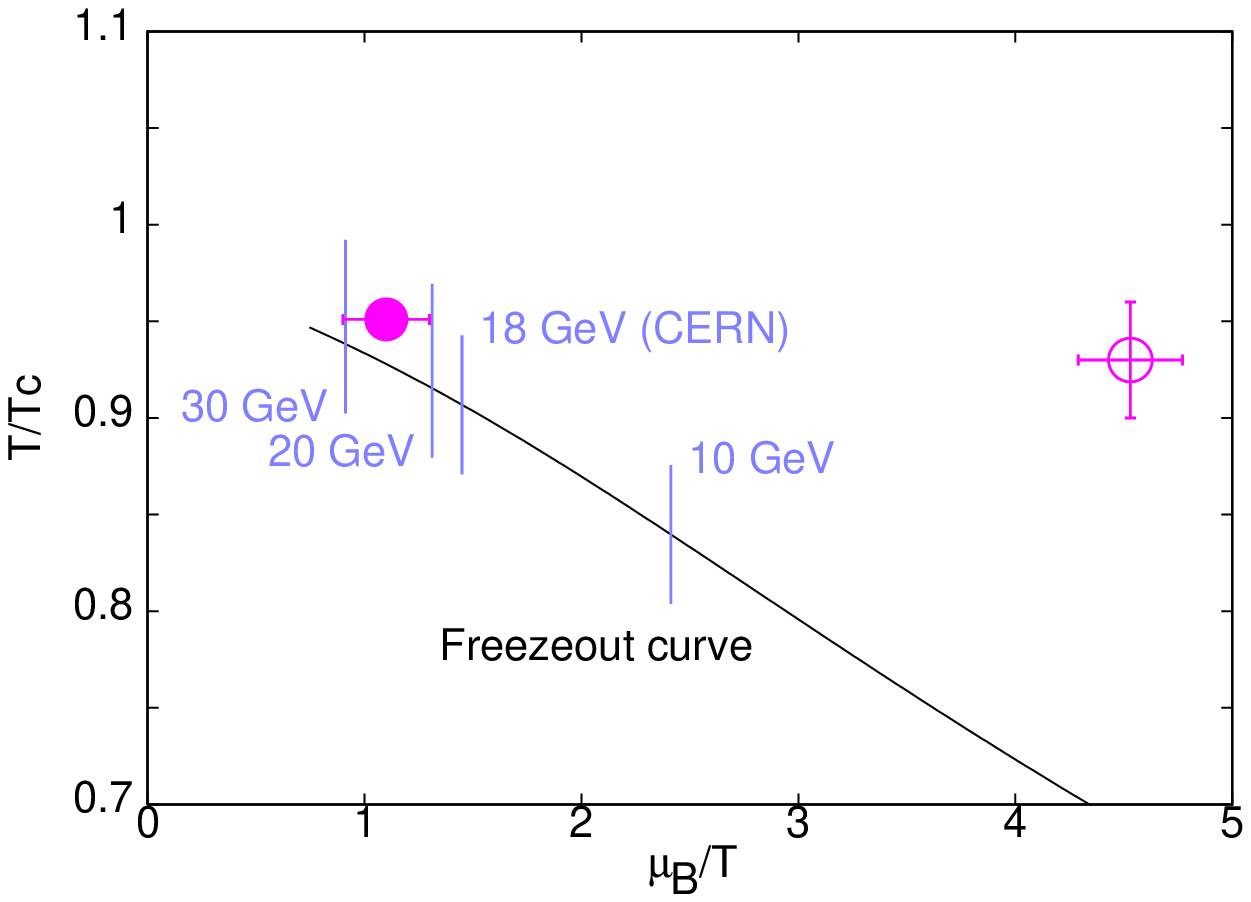}}
\end{minipage}
\vspace{-0.5cm}
 \caption{QCD phase diagram in T vs. $\mu_{B}$ plane showing the QCP from various 
model calculations. For details see text. From left to right the plots are adapted
from Refs.~\cite{kapusta},\cite{fodor3}, and \cite{gupta1}, respectively.} 
 \label{qcpth}
\end{figure}

Several QCD based models predict the existence of an end point at high $\mu_{B}$ 
for the first order phase transition in the QCD phase diagram~\cite{raja}. However the exact
location depends on the model assumptions used. A summary of these model results
can be found in Ref.~\cite{sumQCP}. One such calculation based on linear $\sigma$ 
model coupled to two flavors of identical mass quarks was presented at the conference~\cite{kapusta}.
This work extends the previous investigations by including 
thermal fluctuations of the meson and fermion fields. It is observed that when all other 
parameters of the model are fixed, the calculation allows for existence of zero, one or two
critical points in phase diagram depending on the value of the vacuum pion mass. A typical
result from this calculation for varying mass of pion is shown in Fig.~\ref{qcpth}. 
Given the ambiguity in predictions of QCP in models, studies on lattice 
was expected to provide reliable estimates. 
However lattice calculations at finite $\mu_{B}$ have important issues to be addressed. 
Typically, for any lattice computation one needs to evaluate the
expectation value of an observable $X$,
$\langle X(m_v) \rangle = {{ \int \scriptstyle{D} U \exp (-S_G)
X(m_v)~ {\rm Det}~M(m_s)}
\over { \int \scriptstyle{D} U \exp (-S_G)~ {\rm Det}~M(m_s)} }~~,$
where $M$ is the Dirac matrix in $x$, colour, spin, flavour space for sea
quarks of mass $m_s$, $S_G$ is the gluonic action, and the observable
$X$ may contain fermion propagators of mass $m_v$.
The Det M for non zero $\mu$ is not positive definite, hence numerical methods of evaluation 
of the expectation values is difficult, this is commonly refered to as the {\it sign problem}.
There are several ways suggested to overcome this issue. (i) Reweighting the partition function
in the vicinity of transition temperature and $\mu$ = 0~\cite{fodor3}, (ii) Taylor expansion of thermodynamic
observables in $\mu$/T about $\mu$ = 0~\cite{gupta} and (iii) Choosing the chemical potential to be imaginary
will make the ferminonic determinant positive~\cite{forcrand}. The first two methodologies yield an 
existence of QCP (as shown in Fig.~\ref{qcpth}), whereas the thrid procedure gives a
QCP only when the first co-efficient in the Taylor expansion of generic quark mass on
the chiral critical surface ($m_{c}$) as a function of  
$\mu$/T ($\frac{m_c(\mu)}{m_c(0)} = 1+\sum_{k=1} c_{k} \left(\frac{\mu}{\pi T_c}\right)^{2k}$) 
is positive. The calculations seem to support a negative value of $c_{1}$ 
indicating an absence of QCP\footnote{
Developments after QM2009 have shown that with decreasing
lattice spacing, $c_1$ becomes positive and the expected picture of the phase
diagram seems to be recovered (O. Philipsen at CPOD2009).}
This result was 
intensely discussed at the conference with suggestion to do the calculation with larger
spatial volume, check the stability of the results for different values of $N_{\tau}$ and
for a realistic 2+1 flavour in continum limit. A new calculation on lattice using
the Canonical ensemble was presented at the conference showing the existence of 
an end point for $T_E$ $\sim$ 160 MeV and $\mu_E$ $\sim$ 600 MeV~\cite{li}. Further improvements
are expected when calculations will use a realistic pion mass, instead of $m_\pi$ 
= 700 MeV used in the presented work.

\subsection{Signatures of QCD Critical Point and Experimental Results}

\begin{figure}[tbp]
\begin{minipage}[t]{0.37\textwidth}
\centerline{ \includegraphics[width=1.07\textwidth]{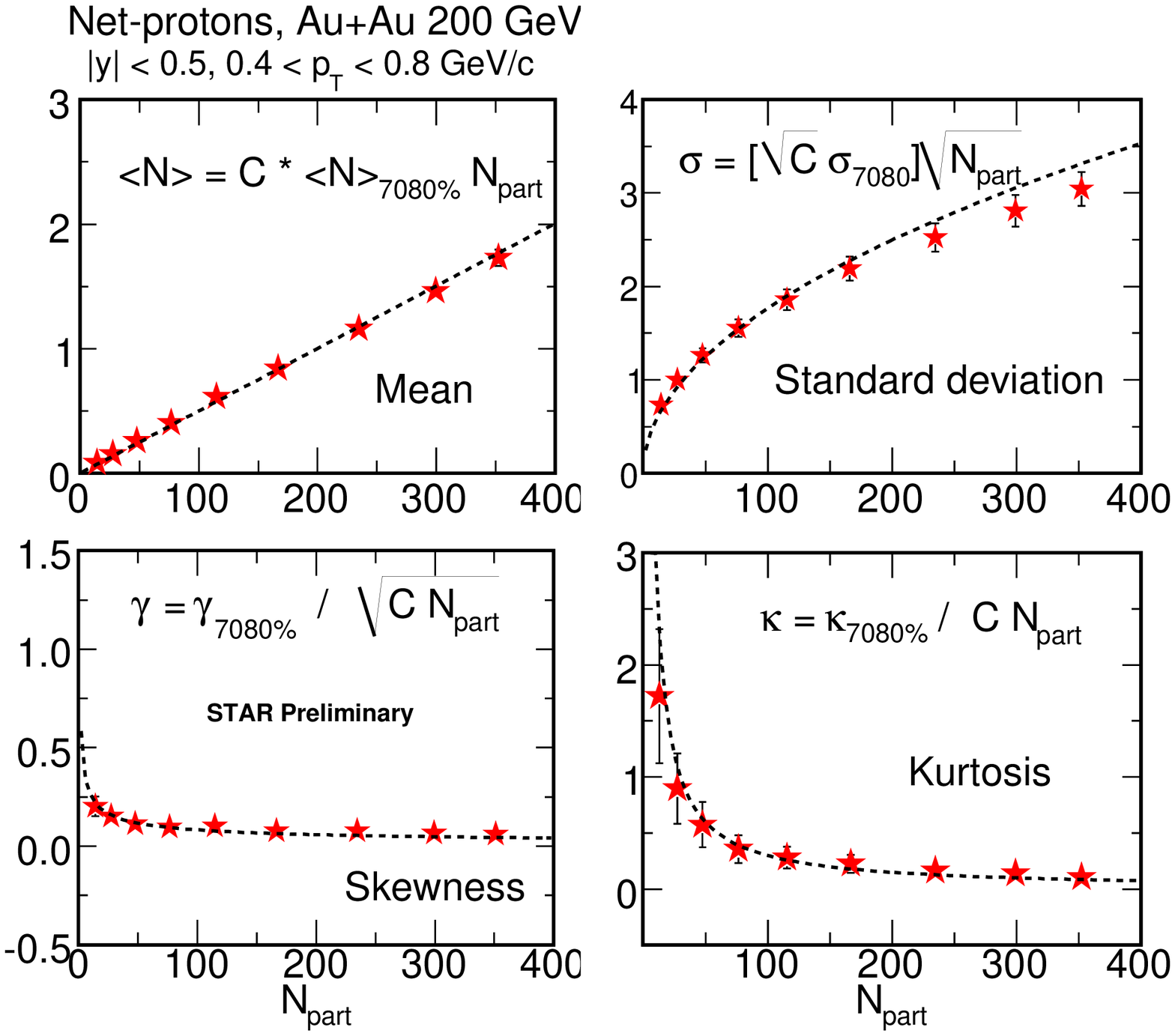}}
\end{minipage}
\begin{minipage}[t]{0.3\textwidth}
\centerline{ \includegraphics[width=1.04\textwidth]{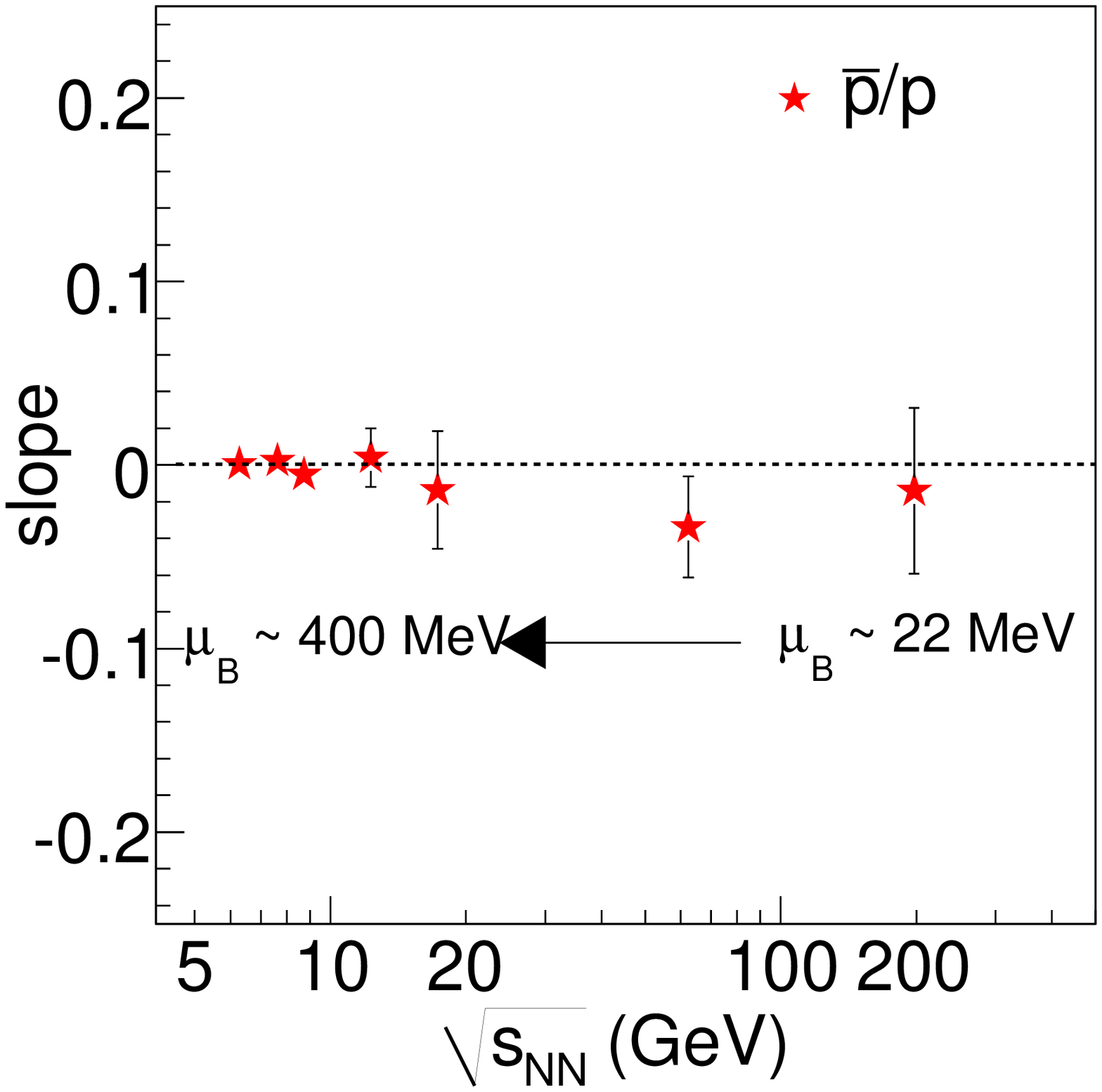}}
\end{minipage}
\begin{minipage}[t]{0.3\textwidth}
\centerline{ \includegraphics[width=1.04\textwidth]{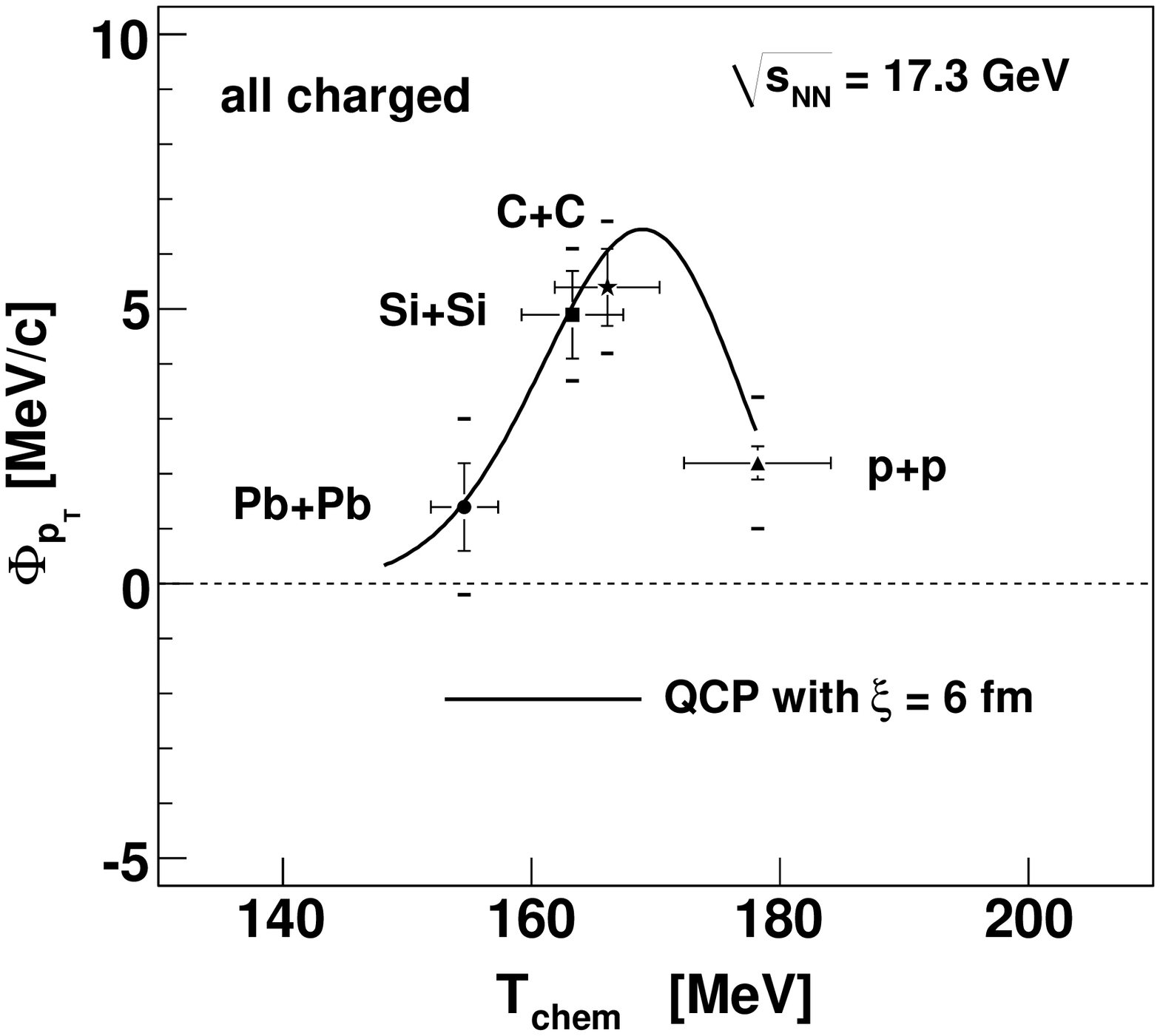}}
\end{minipage}
\vspace{-0.4cm}
 \caption{Moments of net-protons vs. centrality (left panel)~\cite{bm}, slope of anti-baryon to baryon ratio vs. $p_{T}$ (middle panel), system size dependence of mean $p_{T}$ fluctuations (right panel)~\cite{katarina}. See text for details. } 
 \label{qcdex}
\end{figure}

The characteristic signature of QCP is large fluctuations in event-by-event
conserved quantities like net-charge, net-baryon number and net-strangeness~\cite{flucth}. The
variance of these distributions ($\langle (\delta N)^{2} \rangle$) are proportional to square
of the correlation length ($\xi$). The finite size and finite time effects attained 
in high energy heavy-ion collisions, limits the value of the $\xi$ 
achieved in the collisions. Model calculations suggest it could be 
small (2-3 fm)~\cite{corrlen}, thereby making it extremely challenging to measure 
in the experiments. Motivated by the fact that non-Gaussian features in above observables 
increase if the system freezes-out closer to QCP, it has
been suggested to measure higher moments (non-zero skewness and kurtosis 
indicates non-Gaussianity) of net-charge or net-baryon number distributions. 
Further it has been shown that higher moments ($\langle (\delta N)^{3} \rangle$ 
$\sim$ $\xi^{4.5}$ and $\langle (\delta N)^{4} \rangle$ $\sim$ $\xi^{7}$)
have stronger dependence on $\xi$ compared to variance and hence have higher sensitivity~\cite{stephanov}. 
Another important reason to look for higher moments comes from lattice 
calculations of the quadratic and quartic net-charge, net-baryon and net-strangeness 
(these denoted by $N_{X}$) 
fluctuations (evaluated at vanishing chemical potential), 
$\chi_2^X = \frac{1}{VT^3}\langle N_X^2\rangle$ 
and $\chi_4^X = \frac{1}{VT^3}\left(\langle N_X^4\rangle -3 \langle N_X^2\rangle^2\right)$~\cite{cheng1}.
$\chi_2^X$ show a rapid rise in the transition region whereas the  $\chi_4^X$ show
a maximum at $T_{c}$. This maximum is most pronounced for the baryon number 
fluctuations and would diverge at QCP due to long-range correlations~\cite{gupta1}. So studying higher 
moments allows for a connection between QCD calculations on lattice and experimental data. 
Experimentally it is difficult to measure all the produced baryons. This issue was 
addressed theoretically, where it was shown that net-proton number fluctuations would
faithfully reflect the singularity of the charge and  baryon number susceptibility due to the
iso-spin blindness of $\sigma$ field~\cite{stephanov}. Hence looking for non-monotonic variation 
of  higher moments  of net-proton distribution as a function of $\sqrt{s_{NN}}$ (or $T-\mu_{B}$)
is sufficient to locate the QCP. 
The first results on moments of net-proton distribution at RHIC energies was presented at
the conference and shown in Fig.~\ref{qcdex}~\cite{bm}. The evolution of the moments
from preipheral to central collisions follow the expectations (dashed lines) from central limit theorem
(equations shown in Fig.~\ref{qcdex}). 
This study at $\mu_{B}$ $<$ 30 MeV provides an understanding and a formulation for the
physics background for the observable expected to be sensitive to long range fluctuations
as expected for QCP.

Several other interesting signatures of QCP and associated results 
from SPS were discussed at the
conference, one of them was based on the idea that the presence of a critical point 
deforms the isoentropic trajectories in the $T-\mu_{B}$ phase plane. The critical 
point serves as an attractor of the hydrodynamical trajectories. This later 
feature was debated at the conference. If such a scenario exists, it would lead to 
an experimental signal of drop in $\bar{p}/p$ ($\sim$ $e^{-2\mu_{b}/T}$) vs. $p_{T}$ 
(in intermediate region)~\cite{asakawa}. The suggestion was put to test with the existing data, 
by looking at the slope of $\bar{p}/p$ vs. $p_{T}$ at available $\sqrt{s_{NN}}$ 
at SPS and RHIC energies. As shown in the Fig.~\ref{qcdex} (middle panel) there is no large drop in 
slope is observed for intermediate $p_{T}$ range. Another novel idea discussed was based on 
 study of critical dynamics around QCP with relativisic dissipative hydrodynamics~\cite{sound}. Using
the idea of coupling the density fluctuations to thermal energy it was shown that sound
modes around QCP will be suppressed. This will then lead to disappearance of mach-cone
like signals observed at top RHIC energies~\cite{mach}. NA49 experiment presented multiplicity and
mean $p_{T}$ fluctuation results as a function of beam energy and ion size~\cite{katarina}. 
Within the experimental
acceptance no strong non-monotonic dependence of fluctuations was observed for Pb+Pb
collisions as a function of $\sqrt{s_{NN}}$ (or $\mu_{B}$). However the system size
dependence (using the central C+C, Si+Si and Pb+Pb data) seems to indicate a non-monotonic
behaviour around $\mu_{B}$ $\sim$ 250 MeV and $T$ $\sim$ 178 MeV. This was shown to be
consistent with a theory calculation including a QCP and with  $\xi$ $\sim$ 6 fm and folding
in the experimental acceptance (shown as solid line in Fig.~\ref{qcdex}(right)). 
This study further underscores the need for a more detailed
investigation in the upcoming critical point search programs at RHIC and SPS.

\section{Phase Diagram}
\begin{figure}
\centering
\includegraphics[scale=0.37]{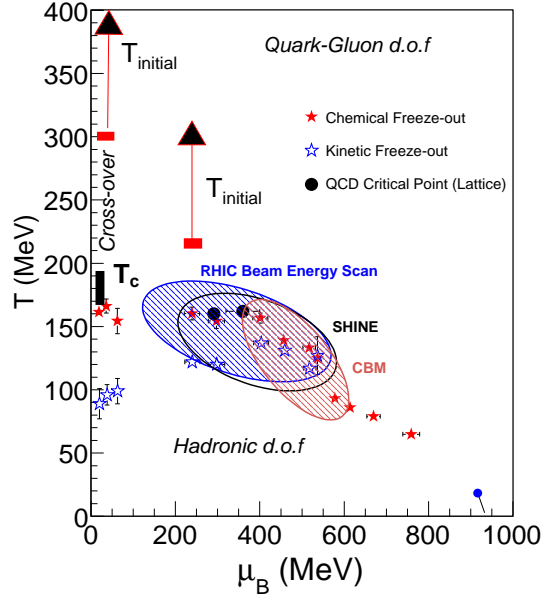}	       
\vspace{-0.6cm}
\caption[]{Current understanding of $T$--$\mu_{B}$ phase diagram for nulear matter from 
theoretical QCD calculations, experimental results and discussions at the conference.
RHIC QCP search: $\sqrt{s_{NN}}$ $\sim$ 39 - 5 GeV (collider); 
SHINE: lighter nuclei, $\sqrt{s_{NN}}$ $\sim$ 17-5 GeV (fixed target) 
and CBM: high luminosity, $\sqrt{s_{NN}}$ $\sim$ 10 - 4 GeV (fixed target).}
\label{phase}
\end{figure}

The current understanding of the QCD phase digram as discussed at the conference 
is depicted in Fig.~\ref{phase}. From the QCD calculations on lattice it is now 
established theoretically that the quark-hadron transition at $\mu_{B}$ = 0 MeV 
is a cross-over~\cite{crossover}. 
The critical temperature for a quark-hadron phase transition lies within a 
range of 170-190 MeV (vertical band in Fig.~\ref{phase})~\cite{rbc,buda}. 
Most calculations on lattice also indicate the existence of 
QCD critical point for $\mu_{B}$ $>$ 160 MeV~\cite{fodor3,gupta1,li}. The exact location is not yet known
unambigiously. Two such predictions computed on lattice are shown in Fig.~\ref{phase}
for a $T_{c}$ of 176 MeV~\cite{fodor3,gupta1}. High energy heavy-ion collision experiments have seen
distinct signatures which suggest that the relevant degrees of freedom at top RHIC~\cite{whitepapers} and 
SPS energies~\cite{claudia} in the initial stages of the collisions are quark and gluons. 
Specifically the direct photon data is used to show 
that the initial temperature ($T_{initial}$, Fig.~\ref{phase}) reached at RHIC and SPS is greater 
than $T_{c}$ predicted by lattice~\cite{directph}. Further the understanding of suppression in high
$p_{T}$ hadron production in heavy-ion collisions relative to $p$+$p$ collisions at RHIC
requires a medium energy density $>$ 1 GeV/fm$^3$ (critical energy density from lattice
for a phase transition)~\cite{whitepapers}. The experiments have also measured the temperature at which 
the inelastic collisions ceases (Chemical freeze-out) and ellastic collisions ceases (Kinetic
freeze-out)~\cite{star}. These temperatures (as shown in Fig.~\ref{phase}) 
are extracted from the measured particle ratios 
and transverse momentum distributions using model calculations which assume 
the system is in chemical and thermal equilibrium. It is interesting to note that the
difference between the two freeze-out temperatures becomes smaller at high $\mu_{B}$
(estimated at chemical freeze-out). 
New experimental programs
at RHIC~\cite{lokesh}, SPS~\cite{shine}, FAIR~\cite{cbm} and 
NICA facilities have been designed to search for the QCD 
critical point in coming years. Whereas the experimental program at LHC 
(probing the region of $\mu_{B}$ $\sim$ 0 MeV of the phase diagram) will 
provide an unique opportunity to understand the properties of matter 
governed by quark-gluon degrees of freedom at unprecedented high initial temperatures 
(higher plasma life time) achieved in the Pb+Pb collisions at 5.5 TeV~\cite{urs}. 
A novel theoretical proposal was made at the conference on the existence of a 
quarkyonic phase around $\mu_{B}$ values corresponding to AGS energies~\cite{larry}. This is 
in addition to confined and de-confined phases. The matter in such a phase is 
expected to have energy density and pressure that of a gas of quarks, and  yet be confined. 
Baryon-Baryon correlations to look for nucleation of baryon rich bubbles surrounded 
by baryon free regions was discussed as a signature of such a phase~\cite{paul}.


\section*{Acknowledgments} 
I would like to thank QM2009 IAC and Organizing committee for inviting me to give this
Rapporteur talk. Also would like to thank Drs. J. Alam, R. Bhalerao, A. Bhasin , 
X. Dong, K. Grebieszkow, S. Gupta, J. M. Heuser, C. Hoehne, J. Kapusta, V. Koch, T. Kunihiro, 
L. Kumar, A. Laszlo, M. Lisa, L. Mclerran, T. K. Nayak, C. Nonaka, P. Petreczky, A. Poskanzer
S. Raniwala, K. Rajagopal, L. Ruan, C. Sasaki, M. Stephanov, S. Voloshin, N. Xu, and Z. Xu 
for many inputs and suggestions. Thanks to RNC group at LBNL for the hospitality during my visit. Financial assistance from the Department of Atomic Energy, Government of India is gratefully acknowledged.


\begin{thebibliography}{00} 
   
\bibitem{tintegral} K. Kanaya et al, these proceedings 
                    and T. Umeda et al., {\it Phys. Rev.} {\bf D 79} (2009) 051501.
\bibitem{action} P. Hegde et al., {\it Eur. Phys. J.} {\bf C 55} (2008) 423.

\bibitem{crossover}Y. Aoki, G. Endrodi, Z. Fodor, S.D. Katz, K.K. Szabo,
                   {\it Nature} {\bf 443} (2006) 675. 

\bibitem{firstorder}S. Ejiri, arXiv:0908.0544
                    and {\it Phys. Rev.} {\bf D 78} (2008) 074507.

\bibitem{rbc}R. Soltz for the HotQCD Collaboration, arXiv:0908.1951 
             and M. Cheng et al., {\it Phys. Rev.} {\bf  D 74} (2006) 054507.

\bibitem{buda}Z. Fodor et al., these proceedings; 
              Y. Aoki et al., {\it JHEP} {\bf 0906} (2009) 088;
              Y. Aoki et al.,  {\it Phys. Lett.} {\bf B 643} (2006) 46.

\bibitem{rgupta}R. Gupta for HotQCD Collaboration, arXiv:0810.1764.

\bibitem{cheng}M. Cheng et al., {\it Phys. Rev.} {\bf D 77} (2008) 014511 
               and P. Petreczky, arXiv:0908.1917.

\bibitem{fodor2}G. Endrodi et al., {\it PoS LAT2007} (2007) 228.

\bibitem{marco} M. Panero, arXiv:0907.3719.

\bibitem{voloshin} S. A. Voloshin for STAR Collaboration, arXiv:0907.2213. 


\bibitem{kharzeev} Dmitri E. Kharzeev et al.,  {\it Nucl. Phys.} {\bf  A803} (2008) 227 
                   and Kenji Fukushima et al., {\it Phys. Rev.} {\bf  D 78} (2008) 074033.

\bibitem{raja} K. Rajagopal and F. Wilczek, arXiv:hep-ph/0011333.

\bibitem{sumQCP} M. Stephanov, {\it Prog. Theor. Phys. Suppl.} {\bf  153} (2004) 139; 
                {\it Int. J. Mod. Phys.} {\bf A 20} (2005) 4387.

\bibitem{kapusta}E.S. Bowman and J.I. Kapusta,
                 {\it Phys. Rev.} {\bf C79} (2009) 015202.

\bibitem{fodor3}Z. Fodor and S.D. Katz, {\it JHEP} {\bf 0404} (2004) 50.

\bibitem{gupta}R. Gavai and S. Gupta, {\it Phys. Rev.} {\bf D 68} (2003) 034506.

\bibitem{forcrand} O. Philipsen, arXiv:0907.4668.

\bibitem{gupta1} R. V. Gavai and S. Gupta, {\it Phys. Rev.} {\bf  D 78} (2008) 114503;
                {\it Phys. Rev.} {\bf D 78} (2008) 14503; {\it Phys. Rev.} {\bf D 71} (2005) 
                 114014.

\bibitem{li} Anyi Li et al., arXiv:0908.1155.

\bibitem{flucth} M. A. Stephanov, K. Rajagopal and E. Shuryak, 
                 {\it Phys. Rev.} {\bf  D 60} (1999) 114028;
                 {\it Phys. Rev. Lett.} {\bf 81} (1998) 4816; 
                 V. Koch, arXiv:0810.2520.

\bibitem{corrlen} B. Berdnikov and K. Rajagopal, {\it Phys. Rev.} {\bf D 61} (2000) 105017.


\bibitem{stephanov} M.A. Stephanov, {\it Phys. Rev. Lett.} {\bf 102} (2009) 032301;
                    Y. Hatta and  M.A. Stephanov, {\it Phys. Rev. Lett.} {\bf 91} (2003) 102003.

\bibitem{cheng1} M. Cheng et al., {\it Phys. Rev.} {\bf D 79} (2009) 074505; 
                 Chuan Miao, for RBC-Bielefeld Collaboration, arXiv:0907.4624 and C. Sasaki, arXiv:0907.4713.





\bibitem{bm} B. Mohanty for the STAR Collaboration (Poster at QM2009); 
             T. K. Nayak for the STAR Collaboration,arXiv:0907.4542.

\bibitem{katarina}K. Grebieszkow, for the NA49 Collaboration, arXiv:0907.4101. 

\bibitem{asakawa} C. Nonaka et al., arXiv:0907.4435 
                  and M. Asakawa et al., {\it Phys. Rev. Lett.} {\bf 101} (2008) 122302.

\bibitem{sound} T. Kunihiro et al., arXiv:0907.3388.

\bibitem{mach} B. I. Abelev et al., STAR Collaboration, 
               {\it Phys. Rev. Lett.} {\bf  102} (2009) 52302.




\bibitem{whitepapers}  BRAHMS Collaboration, I. Arsene et al.,
        {\it Nucl. Phys.} {\bf A 757} (2005) 1;
        PHOBOS Collaboration, B.B. Back et al.,
        {\it Nucl. Phys.} {\bf A 757} (2005) 28;
        STAR Collaboration, J. Adams et al.,
        {\it Nucl. Phys.} {\bf A 757} (2005) 102;
        PHENIX Collaboration, K. Adcox et al.,
        {\it Nucl. Phys.} {\bf A 757} (2005) 184.

\bibitem{claudia} C. Hohne, arXiv:0907.4692.

\bibitem{directph} A. Adare et al (PHENIX Collaboration),arXiv:0804.4168;
                   M. M. Aggarwal et al., (WA98 Collaboration), 
                   Phys. Rev. Lett. {\bf 85} (2000) 3595.

\bibitem{star} B.I. Abelev et al., STAR Collaboration, 
               {\it Phys. Rev.} {\bf  C 79} (2009) 34909 and references therein.

\bibitem{lokesh} L. Kumar for the STAR Collaboration, arXiv:0907.1943;
                 B. I. Abelev et al., STAR Collaboration, arXiv:0909.4131.


\bibitem{shine} A. Laszlo, for the NA61/SHINE Collaboration, arXiv:0907.4493.

\bibitem{cbm}Johann M. Heuser for CBM Collaboration, arXiv:0907.2136.

\bibitem{urs} Urs Achim Wiedemann, arXiv:0908.2294.

\bibitem{larry} L. McLerran and R. D. Pisarski, 
                {\it Nucl. Phys.} {\bf  A 796} (2007) 83; 
                 L. McLerran, arXiv:0907.4489.

\bibitem{paul} A. Mocsy and P. Sorensen, Poster at QM2009.

\end{thebibliography}
\end{document}